\documentclass[preprint,showpacs,preprintnumbers,amsmath,amssymb]{revtex4}
\usepackage{graphicx}
\usepackage{dcolumn}
\usepackage{bm}

\begin{document}


\title{Rashba effect in the graphene/Ni(111) system}

\author{Yu. S. Dedkov,$^{1,}$\footnote{Corresponding author. Electronic address: dedkov@physik.phy.tu-dresden.de} M. Fonin,$^2$ U. R\"udiger,$^2$ and C. Laubschat$^1$}
\affiliation{\mbox{$^1$Institut f\"ur Festk\"orperphysik, Technische Universit\"at Dresden, 01062 Dresden, Germany}\\
$^2$Fachbereich Physik, Universit\"at Konstanz, 78457 Konstanz, Germany}

\date{\today}

\begin{abstract}

Here, we report on angle-resolved photoemission studies of the electronic $\pi$ states of high-quality epitaxial graphene layer on a Ni(111) surface. In this system electron binding energy of the $\pi$ states shows a strong dependence on the magnetization reversal of the Ni film. The observed extraordinary large energy shift up to 225\,meV of the graphene-derived $\pi$-band peak position for opposite magnetization directions is attributed to a manifestation of the Rashba interaction of spin-polarized electrons in the $\pi$ band with the large effective electric field at graphene/Ni interface. Our findings show that an electron spin in the graphene layer can be manipulated in a controlled way and have important implications for graphene-based spintronic devices.

\end{abstract}


\maketitle

The ability to manipulate a single electron spin using an external electric field would be crucial for nanoscopic spintronics~\cite{Wolf:2001,Awshalom:2004,Wolf:2006} and related technological applications where both the charge and spin of electrons are utilized to represent information bits or to perform data processing. For example, the idea of a spin field-effect transistor (spin-FET) device introduced by Datta and Das~\cite{Datta:1990} is based on the possibility to manipulate the spin of the injected electron via application of only an electrical field without any additional magnetic field. As it was realized early on by Rashba~\cite{Rashba:1960} an electric field acts as an additional magnetic field in the rest frame of a moving electron causing an energy splitting or spin precession.

At any crystal interface or surface a structural inversion asymmetry gives rise to the Rashba Hamiltonian: $H_R=\alpha_R(\mathbf{e}\times\mathbf{p})\cdot\mathbf{s}$, where $\mathbf{s}$ is the spin of an electron, moving with momentum $\mathbf{p}=\hbar\mathbf{k}$ in an electric field $\mathbf{e}$ and introduces an effective magnetic field perpendicular to $\mathbf{k}$. The so-called Rashba coefficient, $\alpha_R$, is proportional to the electric field and depends further on the effective, material-dependent, spin-orbit coupling strength. This additional Hamiltonian modifies the energy spectrum of electrons in the valence band and imposes a splitting on the electronic bands which is proportional to the electron wave number $k$: $\Delta E(\mathbf{k})=E(\mathbf{k},+\mathbf{M})-E(\mathbf{k},-\mathbf{M})$. Here, $\pm\mathbf{M}$ denotes two opposite directions of magnetization which is connected with spin of electron, $\mathbf{s}$. To date, this effect has been found on non-magnetic heavy-element surfaces (Au(111)~\cite{LaShell:1996,Hoesch:2004}, Bi/Ag(111)~\cite{Ast:2007}), and magnetic (Gd(0001)~\cite{Krupin:2005}) surfaces. Moreover, Rashba effect has been reported for two-dimensional electron gases in semiconductor heterostructures where a zero-magnetic-field splitting was observed~\cite{Zutic:2004} with a more recent demonstration of a gate-voltage control of the Rashba interaction in some semiconductor structures~\cite{Sih:2007}. However, a high efficient spin-FET device (see Fig.\,1,a) requires long spin relaxation time as compared to the mean time of transport through the channel combined with a sufficient difference of the spin rotation angles between two states ("0" and "1") as well as an insensitivity of spin rotation to the carrier energy. In the recent theoretical work~\cite{Semenov:2007} spin-FET was proposed on the basis of a single graphene layer, a novel material consisting of a flat monolayer of carbon atoms packed in a two-dimensional honeycomb-lattice, in which the electron dynamics is governed by the Dirac equation~\cite{Wilson:2006,Geim:2007}. Long electronic mean paths~\cite{Geim:2007} and negligible spin-orbit coupling of the carbon based system~\cite{Min:2006}, i.e. large spin relaxation times, make graphene a best-choice material for the observation of near ballistic spin transport.

Here we report a strong Rashba ``splitting'' for the $\pi$-states in the epitaxial graphene layer on the Ni(111) surface. Instead of a large spin-orbit interaction like in heavy atom materials we make use of the effect of a strong effective electric field ($\mathbf{E}$) due to a large asymmetry of a charge distribution at the graphene/Ni(111) interface caused by hybridization between the $\pi$ states of graphene with the Ni $3d$ states~\cite{Nagashima:1994,Dedkov:2001}. On the other hand, strong hybridization of the graphene $\pi$-states with the spin-split Ni $3d$ states in this system leads to the fact that the $\pi$ states become spin-polarized and thus the sign of magnetization of the graphene/Ni(111) system can be reversed by applying a magnetic field of opposite direction. The experimental geometry is shown in Fig.\,1,b. The graphene/Ni(111) sample is illuminated by the ultra-violet light (He\,II$\alpha$, $h\nu=40.8$\,eV) and angle-resolved photoemission spectra are measured for two opposite magnetization directions of the sample. The Rashba effect can be observed in this geometry due to the fact that three vectors ($\mathbf{E}$, $\mathbf{k}$, $\mathbf{s}$) are orthogonal to each other. The parasitic contributions of magnetic linear dichroism (MLD) are expected to be very small due to the weak spin-orbit interaction in graphene and may be effectively supressed in this geometry by placing the sample perpendicular to the incident light. Additionally, the Rashba effect results in a linear $\mathbf{k}$-dependent energy shift of the electronic band dispersion whereas MLD would lead to a variation of the intensities of photoemission peaks which makes it possible to distinguish between the two effects.

Experiments were performed in two separated experimental stations in equal conditions. Scanning tunneling microscopy measurements were performed in ultrahigh vacuum (UHV) with an Omicron VT AFM/STM at room temperature using electrochemically etched tungsten tips that were flash annealed by electron bombardment. The ``$\pm$'' sign of the bias voltage denotes the voltage applied to the sample. Photoelectron spectra were recorded at room temperature in angle-resolved mode with a $180^\circ$ hemispherical energy analyzer SPECS PHOIBOS 150. The energy resolution of the analyzer was set to 50\,meV and the angular resolution was $0.5^\circ$. The light incident angle was 20$^\circ$ with respect to the sample surface. Angle-resolved measurements were performed in magnetic remanence after having applied a magnetic field pulse ($\pm\mathbf{M}$) of about 500\,Oe along the $<1\bar{1}0>$ easy magnetization axis of the Ni(111) thin film. A well ordered Ni(111) surface was prepared by the thermal deposition of Ni films with a thickness of about 150\,\AA\ on to the clean W(110) substrate followed by a subsequent annealing at 300$^\circ$\,C. Prior to the film preparation the tungsten substrate was cleaned by several cycles of oxygen treatment and subsequent flashes at about 2300$^\circ$\,C. An ordered graphene overlayer on Ni(111) was prepared via cracking of propene gas (C$_3$H$_6$) according to a recipe described in Ref.~\cite{Dedkov:2001}. A LEED study of the graphene/Ni(111) system reveals a well-ordered $p(1\times1)$-overstructure as expected from the small lattice mismatch of only 1.3\%. STM measurements indicate that a continuous epitaxial graphene layer was formed on the Ni(111) substrate. 

Recent photoemission experiments on qusiparticle dynamics in graphene have been performed either on graphene layers formed on top of SiC(0001) substrates~\cite{Otah:2006,Bostwick:2007a,Bostwick:2007b} or on single crystalline graphite~\cite{Zhou:2006}. Here we prepare a high quality epitaxial graphene layer via cracking of propene gas (C$_3$H$_6$) on top of a Ni(111) surface~\cite{Nagashima:1994,Dedkov:2001}. After the cracking procedure the Ni(111) surface is completely covered by the graphene film as demonstrated by scanning tunneling microscopy (STM). An STM image of the graphitized Ni(111) surface, acquired at room temperature and at a sample bias of +0.05\,V, is shown in Fig.\,2,a. All investigated terraces display the same atomic structure showing three different levels of apparent heights (Fig.\,2,b). Section of \textbf{b} is overlaid by the honeycomb lattice of graphene. The distance between the two atoms within A or B graphene sublattices (bright maxima or dark minima) was measured to be 2.4$\pm$0.1\,\AA\, being in good agreement with the expected interatomic spacings of 2.46\,\AA\ in graphene. Investigations by means of angle-resolved photoemission are in excellent agreement with previous studies~\cite{Nagashima:1994,Dedkov:2001} and demonstrate the presence of a single graphene layer on the Ni(111) surface.  Figure\,2,c shows photoemission spectra taken with He\,II$\alpha$ radiation ($h\nu=40.8$\,eV) along the $\overline{\Gamma}-\overline{M}$ direction of the surface Brillouin zone (inset of Fig.\,2,c). For two dimensional systems, like graphene, the energy dispersion can be described as a function of only the in-plane component of the wave vector, $\mathbf{k_{||}}$. In the following, describing $\pi$ states of graphene we will not distinguish between $\mathbf{k}$ and $\mathbf{k_{||}}$. From a comparison of the photoemission spectra of graphene/Ni(111) with pure graphite~\cite{Shikin:1999} we conclude that the difference in the binding energy of the $\pi$-states is about 2.3\,eV which is close to the value observed earlier~\cite{Nagashima:1994,Dedkov:2001} and in good agreement with the theoretical prediction of 2.35\,eV~\cite{Bertoni:2005}. This shift reflects the effect of hybridization of the graphene $\pi$ bands with the Ni $3d$ bands and to less-extend with Ni $4s$ and $4p$ states so that a charge density gradient arise in the interface layer.

In Fig.\,3,a we show two representative pairs of photoemission spectra measured for two opposite magnetization directions ($\pm\mathbf{M}$) and for the emission angles $\theta=0^\circ$ and $\theta=24^\circ$ corresponding to the wave vectors $k=0$\,\AA$^{-1}$ and $k=1.16$\,\AA$^{-1}$, respectively. A clear shift of $\Delta E\approx200$\,meV in the peak position of the $\pi$ states upon magnetization reversal is observed for spectra taken in off-normal emission geometry whereas no shift is observed for spectra measured at normal emission. Although the states for opposite magnetization directions ($\pm\mathbf{M}$) do not exist simultaneously (Fig.\,3,a), the observed energy shift is equivalent to the Rashba splitting at a nonmagnetic surface, and we shall further refer to it as the Rashba "splitting". The energy dispersions extracted from the photoemission spectra measured for two opposite directions of magnetization are shown in Fig.\,3,b together with a zoom in the region around $\overline{\Gamma}$ point. Due to the additional Rashba energy both dispersions are shifted with respect to $\mathbf{k}=0$ with total $\Delta k=0.08\pm0.02$\,\AA. In Fig.\,3,c we plot the magnitude of the Rashba ``splitting'' $\Delta E(\mathbf{k})$ obtained from the data shown in Fig.\,3,b as a function of the wave vector $k$ along the $\overline{\Gamma}-\overline{M}$ direction of the surface Brillouin zone. The linear dependence of $\Delta E$ on $k$ is a clear proof of the Rashba effect in the graphene layer on top of Ni(111) surface. The Rashba ``splitting'' disappears upon approaching the $\overline{M}$ point at the border of the surface Brillouin zone. This can be explained due to the fact that the branches have to exchange their energy positions by crossing the border of the Brillouin zone. The same effect is expected at all other directions. The experimental proof for the other high-symmetry direction, $\overline{\Gamma}-\overline{K}$, is not possible in this experimental geometry since sample is magnetized along the $<1\bar{1}0>$ easy axis of magnetization of the Ni(111) substrare which is parallel to $\overline{\Gamma}-\overline{K}$. For practical use of the effect all energy branches should be pushed to lower binding energies via hole-doping (for example, iodine or FeCl$_3$~\cite{Dresselhaus:1981}) with a sizable potential gradient in order to produce the detectable Rashba ``splitting''. The weak MLD effects visible at $\theta=24^\circ$ for the Ni $3d$ states leads only to changes in photoemission intensities and can not be responsible for the energy shift of the band. Moreover, as mentioned earlier, such dichroism effects can not be detected for the $\pi$ states of graphene because of  an extremely weak spin-orbit interaction in carbon.

The large Rashba ``splitting'' of the $\pi$ states observed in this experiment is thus most probably due to the asymmetry of the charge distribution at the graphene/Ni(111) interface. The hybridization between the $\pi$ states of graphene and the Ni $3d$ states leads to a considerable charge transfer from $3d$ states of nickel surface atoms to the unoccupied $\pi^*$ states of the graphene monolayer~\cite{Yamamoto:1992}. As a consequence, a large potential gradient is formed at the interface and the electrons in the interface state are subject to an effective spatially averaged crystal electric field that leads to the Rashba effect. A similar effect was recently observed at the oxygen/Gd(0001) interface~\cite{Krupin:2005}. There, the Rashba effect at the clean Gd(0001) surface is relatively small and can be explained mainly as originating from spin-orbit interaction in heavy-atom Gd. After adsorption of oxygen a Gd-oxide is formed and a large charge transfer from the subsurface Gd layer to the oxide layer was observed. \textit{Ab initio} calculations without inclusion of spin-orbit interaction demonstrate that in this case the gradient of potential plays the crucial role as origin of the Rashba ``splitting''. However, this model can be implemented for the graphene/Ni(111) system only in case of the spin-polarized $\pi$ band of graphene. The spin-polarized carriers in the graphene layer can be due to hybridization or the proximity effect, which for example was observed for the Al/EuO interface~\cite{Tedrow:1986} (EuO is ferromagnetic insulator and Al is close in electronic configuration of the valence band to carbon). Also, in C/Fe multilayers~\cite{Mertins:2004} and carbon nanotubes on Co substrate~\cite{Cespedes:2004} magnetic moments of carbon atoms in order of 0.05-0.1\,$\mu_B$ were found at room temperature. On the other hand, since the Fermi surface of graphene is centered around the $\overline{K}$ point in reciprocal space where the Ni substrate has states of pure minority spin character~\cite{Karpan:2007} one may expect that predominantly spin-down electrons will fill $\pi^*$ states and produce the observed spin polarization of these states. We conclude that the observed Rashba effect in this system is due to interaction of spin-polarized electrons in the $\pi$ band of the graphene layer with the large effective electric field at graphene/Ni(111) interface.

In conclusion, we report the first experimental evidence of a large Rashba ``splitting'' in an epitaxial graphene layer on top of a Ni(111) substrate. By means of angle-resolved photoelectron spectroscopy we detect a large energy shift of up to 225\,meV of the graphene-derived $\pi$ states upon magnetization reversal of the Ni(111) film. The observed Rashba effect is attributed to the combination of two effects: spin-polarization of the $\pi$-band and effective potential gradient appearing at the graphene/Ni interface. We propose that this effect may be tailored by hole-doping of the graphene layer or by intercalation of another material underneath the graphene layer (for example, Cu~\cite{Dedkov:2001}) leading to a change of interaction in the system. Our findings indicate that a direct control on the electron spin can be obtained via application of an external electric field. The present results are of crucial importance for the design of new carbon-based spintronic devices, particularly at ferromagnet/nonmagnet interfaces where Rashba and exchange interaction together control the electron spin.

This work was funded by the Deutsche Forschungsgemeinschaft (DFG) through Collaborative Research Centers SFB-463 (Projects B4 and B16) and SFB-513 (Project B14).

\newpage

\textbf{Figure captions:}
\newline
\newline
\textbf{Fig.\,1.}\,\,(Color online) Proposed graphene-based experiments. \textbf{a,} Schematic illustration of a graphene-based spin FET device. Spin-polarized electrons are injected from the source FM1 and detected by the drain FM2. Spin-manipulation is possible by applying of bias voltage, V, to the gate separated from graphene layer by an insulator. \textbf{b,} Geometry of the photoemission experiment discussed in the letter. Due to the fact that three vectors (\textbf{E, k, s}) are orthogonal to each other the Rashba effect can be observed in such geometry. 
\newline
\newline
\textbf{Fig.\,2.}\,\,(Color online) High-quality graphene/Ni(111) system. \textbf{a,} Constant current STM image of the graphene/Ni(111) surface (+0.05\,V). The inset shows a LEED image obtained at 63 eV. \textbf{b,} Constant current image showing atomic structure of the graphene monolayer (-0.04\,V). Part of \textbf{b} is superimposed by the honeycomb lattice of graphene. \textbf{c,} Angle-resolved photoemission spectra measured with $h\nu=40.8$\,eV along $\overline{\Gamma}-\overline{M}$ direction of the surface Brillouin zone. Ni $3d$ as well as $\pi$ and $\sigma$ states of graphene layer are marked in the figure.  
\newline
\newline
\textbf{Fig.\,3.}\,\,(Color online) Manifestation of the Rashba effect in graphene layer on Ni(111). \textbf{a,} Series of representative photoelectron spectra of the graphene/Ni(111) system for two different emission angles (marked in the plot) and two directions of magnetization, respectively (according to Fig.\,1,b). A clear energy shift of $\pi$ states is visible for spectra taken at $\theta=24^\circ$. \textbf{b,} Energy dispersions extracted from the photoemission spectra measured for two opposite directions of magnetization. \textbf{c,} Rashba ``splitting'' $\Delta E(\mathbf{k})$ obtained from the data shown in \textbf{b} as a function of the wave vector $k$ along the $\overline{\Gamma}-\overline{M}$ direction of the surface Brillouin zone.    

\clearpage
\begin{figure}[t]\center
\vspace{1cm}
\large \textbf{Fig.\,1, Yu. S. Dedkov \textit{et al.}, see Fig1.jpg}
\end{figure}

\clearpage
\begin{figure}[t]\center
\vspace{1cm}
\large \textbf{Fig.\,2, Yu. S. Dedkov \textit{et al.}, see Fig2.jpg}
\end{figure}

\clearpage
\begin{figure}[t]\center
\vspace{1cm}
\large \textbf{Fig.\,3, Yu. S. Dedkov \textit{et al.}, see Fig3.jpg}
\end{figure}

\end{document}